\documentclass[aps,prl,twocolumn,superscriptaddress,showpacs]{revtex4}
\usepackage{amsmath}
\usepackage{graphicx,amssymb}
\usepackage{indentfirst,latexsym,bm}
\begin{document}
\title{Effect of decoherence on the Berry phase of a spin-half in a rotating magnetic field }
\author{Xin Li}
\affiliation{Department of Physics and State Key Laboratory of Surface Physics, Fudan University, Shanghai, 200433, China}
\affiliation{Faculty of science, Kunming University of Science and Technology, Kunming 650093, China}
\author{Yu Shi}
\email{yushi@fudan.edu.cn}
\affiliation{Department of Physics and State Key Laboratory of Surface Physics, Fudan University, Shanghai, 200433, China}

\begin{abstract}
We investigate the decoherence effect of a bosonic bath  on the Berry phase of a spin-$\frac{1}{2}$ in a time-dependent magnetic field,  without making the Markovian approximation. A two-cycle process resulting in a pure Berry phase is considered. The low-frequency quantum noise significantly affects the Berry phase. In the adiabatic limit, the high-frequency quantum noise only has a small effect.   The result is also valid in some more general situations.
\end{abstract}
\maketitle

\section{Introduction}

Geometric phases~\cite{berry} have been used as an approach to fault-tolerant quantum computing due to its global geometric feature~\cite{rasetti,computing,shi}. How the geometric phase is really insensitive to classical and quantum noises becomes an interesting topic subject to experimental investigations~\cite{leek,filipp,berger}. Theoretically, insensitivity to classical control noise in some circumstances has been demonstrated~\cite{classical}. The effect of quantum noise or decoherence has also been studied,   most of which were for Markovian dynamics~\cite{mark,whitney}, but
Non-Markovian dynamics has also been considered to some extent~\cite{nonmark}.  Also related is  the subject of geometric phases of mixed states~\cite{11}.

In this paper we make a general analysis on the effect of decoherence with non-Markovian dynamics on the Berry phase of spin$-\frac{1}{2}$ coupled to a magnetic field, a set-up which is useful for quantum computing~\cite{shi}. The Markovian limit is also discussed.  Using a master equation approach without making Markovian approximation, and avoiding defining the Berry phase for a mixed state, we calculate the degrading of the fidelity due to the coupling with an environment.

\section{The model and the master equation \label{model}}

Consider a spin-$\frac{1}{2}$ coupled to a rotating magnetic field and to an environment, which is a bosonic bath. The total Hamiltonian is
\begin{equation}
H(t) = H_s+V+H_e \label{eq1}
\end{equation}
with
\begin{eqnarray}
H_s(t)&=&\frac{1}{2}\mathbf{B}(t)\cdot\mathbf{\sigma},\\
V&=& \sigma_{z}
\sum\limits_{k}g_{k}(a_{k}^{\dag}+a_{k}),\\
H_e&=&\sum\limits_{k}\omega_{k}a_{k}^{\dag}a_{k},
\end{eqnarray}
where $\mathbf{B}(t) \equiv (B_{x},B_{y},B_{z})=B(\sin\theta\cos\Omega_0t,\sin\theta\sin\Omega_0t,
\cos\theta)$ is the external field, which rotates around $z$ axis with the angular frequency $\Omega_0$, $\theta$ is the angle between   $\mathbf{B}(t) $ and $z$ axis,  $\mathbf{\sigma} \equiv (\sigma_{x},\sigma_{y},\sigma_{z})$ are the Pauli matrices, $a_{k}^{\dag}$ and $a_{k}$ are the creation and annihilation operators of the field mode $k$, $g_{k}$ is the coupling strength. In an experiment testing the Berry phase of a charge Josephson qubit \cite{leek}, $B_{\|}\equiv\sqrt{B_{x}^{2}+B_{y}^{2}}$ is realized by the dipole interaction strength between the qubit and a microwave field with phase $\Omega_{0}t$, while $B_{z}$ is realized by  the detuning between the qubit transition frequency and the applied microwave frequency.  The time-dependent  Hamiltonian can be transformed to a time-independent one in the rotating frame by a unitary transformation  $U_{1}(t) \equiv e^{\frac{i\sigma_{z}  }{2}\Omega_0t }$~\cite{shi}. Subsequently, the magnetic field can be transformed to be along the $z$ axis in a rotated  frame by another unitary transformation  $U_{2} \equiv e^{\frac{i \sigma_{y} }{2}\alpha }$. With $|\psi\rangle = U_1^{\dagger} U_2^{\dagger} |\tilde{\psi}\rangle$,  the Schr\"{o}dinger equation $
i\partial_t |\psi(t)\rangle = H(t) |\psi(t)\rangle$  is transformed to
\begin{equation}
i\partial_t |\tilde{\psi}(t)\rangle = \tilde{H} |\tilde{\psi}(t)\rangle, 
\end{equation}
where
$
\tilde{H} =U_{2}[U_{1}(t)H(t)U_{1}^{\dag}(t)-iU_{1}(t)\dot{U}_{1}^{\dag}(t)]
U_{2}^{\dag} = \tilde{H}_{s}+\tilde{V}+\tilde{H}_{e}, $
with $\tilde{H}_{s}=\frac{E}{2}\sigma_{z},$ $\tilde{V}=(\cos\alpha\sigma_{z}-\sin\alpha\sigma_{x})\sum\limits_{k}g_{k}(a_{k}^{\dag}+a_{k}),$ $\tilde{H}_{e}=\sum\limits_{k}\omega_{k}a_{k}^{\dag}a_{k},$
where  $\tan\alpha \equiv \frac{B\sin\theta}{B\cos\theta-\Omega_0}$, $E \equiv \sqrt{(B\sin\theta)^{2}+(B\cos\theta-\Omega_0)^{2}}$, which equals the energy gap  between the ground and excited states.

The  density matrix of the composite system consisting of both the spin and the environment obeys the Liouville  equation,
which, in the interaction picture, is
\begin{equation}
\frac{\partial\tilde{\rho}_I(t)}{\partial t} =-i[\tilde{V}_{I}(t),\tilde{\rho}_I(t)],
\end{equation}
with $\tilde{\rho}_I(t)=e^{i(\tilde{H}_{s}+\tilde{H}_{e})t}\tilde{\rho}(t)
e^{-i(\tilde{H}_{s}+\tilde{H}_{e})t}$ and $\tilde{V}_{I}(t)=e^{i(\tilde{H}_{s}+\tilde{H}_{e})t}
\tilde{V}e^{-i(\tilde{H}_{s}+\tilde{H}_{e})t}$. As usual, it is assumed that the initial state is a direct product of the states of the spin and the environment, i.e.,  $\tilde{\rho}_I(0)=\tilde{\rho}^{Is}(0)\otimes\tilde{\rho}^{Ie}(0)$.   The coupling of the bath with the system is weak,  hence $\tilde{\rho}^{Ie}(t)\simeq \tilde{\rho}^{Ie}(0)\equiv\rho^e$.  By applying the projection operator method~\cite{breuer}, one obtains the master equation of the reduced density matrix of the spin $\tilde{\rho}^{Is}(t)$, up to second order of the spin-environment coupling, $
\frac{\partial\tilde{\rho}^{Is}(t)}{\partial t}  = -\int_{0}^{t}dsTr_{E}[\tilde{V}_{I}(t),
[\tilde{V}_{I}(s),\tilde{\rho}^{Is}(t)\otimes\rho^e]]
= -\int_{0}^{t}ds[\langle \epsilon(t)\epsilon(s)\rangle(\sigma(t)\sigma(s)\tilde{\rho}^{Is}(t)-\sigma(s)\tilde{\rho}^{Is}(t)\sigma(t))
+\langle \epsilon(s)\epsilon(t)\rangle(\tilde{\rho}^{Is}(t)
\sigma(s)\sigma(t)-\sigma(t)\tilde{\rho}^{Is}(t)\sigma(s))]  = -\int_{0}^{t}ds[\langle \epsilon(s)\epsilon(0)\rangle(\sigma(t)\sigma(t-s)
\tilde{\rho}^{Is}(t)-\sigma(t-s)\tilde{\rho}^{Is}(t)
\sigma(t))  + \langle \epsilon(0)\epsilon(s)\rangle(\tilde{\rho}^{Is}(t)\sigma(t-s)\sigma(t)-\sigma(t)\tilde{\rho}^{Is}(t)\sigma(t-s))],$
where $\sigma(t)\equiv e^{\frac{i\sigma_{z}}{2}Et}
(\cos\alpha\sigma_{z}-\sin\alpha\sigma_{x})e^{-\frac{i\sigma_{z}}{2}Et}$, $\epsilon(t)\equiv\sum\limits_{k}g_{k}(a_{k}^{\dag}e^{i\omega_{k}t}+
a_{k}e^{-i\omega_{k}t})$,  and $Tr_{E}$ means partial trace over the environment, which has been assumed to be initially in thermal equilibrium at temperature $T$, i.e., $\rho^e= (1-e^{-\omega_{k}/T}) \prod\limits_{k}e^{-\omega_{k}a_{k}^{\dag}a_{k}/T} $.
Our calculation is based on the above master equation, without making the Markovian approximation, i.e., replacing the upper limit of the time integral as infinity, as in some previous studies.

The effect of the environment can be captured by the force autocorrelation function
$\langle \epsilon(t)\epsilon(s)\rangle  = Tr_{E}[\epsilon(t)\epsilon(s)\rho_{e}] = \sum\limits_{k} g_{k}^{2}[N(\omega_k) e^{i(t-s)\omega_{k}}+(N(\omega_k)+1)e^{i(s-t)\omega_{k}}] =  \int_{0}^{\infty}d\omega[\sum\limits_{k}g^{2}_{k}\delta(\omega-\omega_{k})][N(\omega) e^{i(t-s)\omega}+(N(\omega)+1)e^{i(s-t)\omega}] =  \int_{0}^{\infty}d\omega J(\omega)[2N(\omega)\cos((t-s)\omega)+e^{i(s-t)\omega}],$
where $N(\omega_k)
\equiv (e^{\omega_{k}/T}-1)^{-1}$ denotes the average number of bosons in a mode with frequency $\omega_{k}$, $J(\omega)$ is the environment spectrum.

We solve the master equation  by  using  a secular approximation~\cite{breuer} to remove high-frequency terms such as  $e^{inEt}$,  $n$ being an integer, on the right-hand side. Then we obtain the solution of the reduced density matrix $\tilde{\rho}^{Is}(t)\equiv\left[
 \begin{array}{cc}
 \tilde{\rho}^{Is}_{gg}(t) & \tilde{\rho}^{Is}_{ge}(t) \\
\tilde{\rho}^{Is}_{eg}(t) & \tilde{\rho}^{Is}_{ee}(t) \\
\end{array}                                                                                                                         \right]
$, whose matrix elements are
$
\tilde{\rho}^{Is}_{gg}(t) = e^{-n(t)}(m(t)+\tilde{\rho}^{s}_{gg}(0)),$ $
\tilde{\rho}^{Is}_{ge}(t)= e^{-l(t)-ik(t)}\tilde{\rho}^{s}_{ge}(0),$ $
\tilde{\rho}^{Is}_{ee}(t)= 1-\tilde{\rho}^{Is}_{gg}(t),$ $
\tilde{\rho}^{Is}_{eg}(t)= \tilde{\rho}^{Is^{*}}_{ge}(t),$
where $n(t)=4\sin^{2}\alpha\int_{0}^{t} dt_1\int_{0}^{t_{1}} dt_2 \cos(Bt_{2})Re[\kappa(t_{2})],$
$m(t)=2\sin^{2}\alpha\int_{0}^{t} dt_1\int_{0}^{t_{1}} dt_2 Re[\tilde{\kappa}(t_{2})]e^{n(t_{1})},$
$l(t)=\int_{0}^{t}dt_1\int_{0}^{t_{1}} dt_2 (4\cos^{2}\alpha+2\cos(Bt_{2})\sin^{2}\alpha)Re[\kappa(t_{2})],$ $
k(t)=2\sin^{2}\alpha\int_{0}^{t} dt_1 \int_{0}^{t_{1}} dt_2\sin(Bt_{2})Re[\kappa(t_{2})],$
where the energy $E$ has been approximated as $B$, $\kappa(s)\equiv\langle \epsilon(s)\epsilon(0)\rangle$ and $\tilde{\kappa}(s)\equiv e^{-iBs}\kappa(s)$.  It  indicates that the interaction with the environment induces dephasing, energy dissipation and Lamb-like shift, all of which affect the Berry phase to different degrees.
Discussions below will be made in the original frame by using
\begin{equation}
\rho^{s}(t)=U_{1}^{\dag}(t)U^{\dag}_{2}e^{-\frac{i \sigma_{z}}{2}  Et }
\tilde{\rho}^{Is}(t)e^{ i\frac{ \sigma_{z}}{2}  Et }U_{2}U_{1}(t).
\end{equation}

\section{Fidelity \label{fidel} }

To characterize the environmentally induced decoherence, we compare the density matrix  $\rho^{s}(t)$ of the system coupled with the environment, i.e. evolving under $H(t)$,  with  the density matrix $\rho^{s}_0(t)$  of the isolated system,  i.e., evolving under  $H_{s}(t)$. This is done by using the fidelity defined as
\begin{equation}
F(t)=Tr[\rho^{s}(t)\rho^{s}_0(t)],
\end{equation}
by setting $\rho^{s}(0)=\rho^{s}_0(0)$.
Without decoherence, we would have $F(t)=1$. With decoherence, we have $F(t) <1$.
We study how the fidelity $F(t)$, as a function of time, is affected by different environmental spectrum.

The initial state $\rho^{s}(0)$ is set to be an equal superposition of ground and excited states, as  in recent experiments~\cite{leek},  thus    $\rho^s(0)=|\varphi_{0}\rangle\langle\varphi_{0}|$, where $|\varphi_{0}\rangle=\frac{1}{\sqrt{2}}(|e(0)\rangle+|g(0)\rangle)$, with
$|e(t)\rangle=\cos\frac{\theta}{2}|0\rangle+\sin\frac{\theta}{2}e^{i\Omega_0t}|1\rangle$ and  $|g(t)\rangle=\sin\frac{\theta}{2}e^{-i\Omega_0t}|0\rangle-\cos\frac{\theta}{2}|1\rangle$ being instantaneous eigenstates of  $H_{s}(t)$.

The fidelity is calculated to be exactly
$
F(t)=\frac{1}{2}[1+e^{-l(t)}\cos k(t)\cos^{2}\zeta+\sin\zeta+e^{-n(t)}\sin\zeta(\sin\zeta-1-2m(t))],$
where $\zeta\equiv\alpha-\theta$.

We focus on  Ohmic spectrum $J(\omega)=\frac{\lambda}{2}\omega e^{-\frac{\omega}{\Omega}}$, where $\lambda$ is a coupling constant and $\Omega$ is the cut-off frequency. If the correlation time scale of the noise $\tau_{c}=1/\Omega$ is comparable with the adiabatic time scale of the system $\tau_{0}=1/\Omega_0$, which is also the time scale of the dynamics described by the master equation, then the environmental memory affects the system dynamics,  which is then non-Markovian. Hence the Markovian case is defined by $\Omega \gg \Omega_0$, which implies that the short-term dynamical fluctuation of the system caused by the  feedback of the environment is averaged out~\cite{nonm}.

\begin{figure}
\scalebox{0.7}[0.7]{\includegraphics{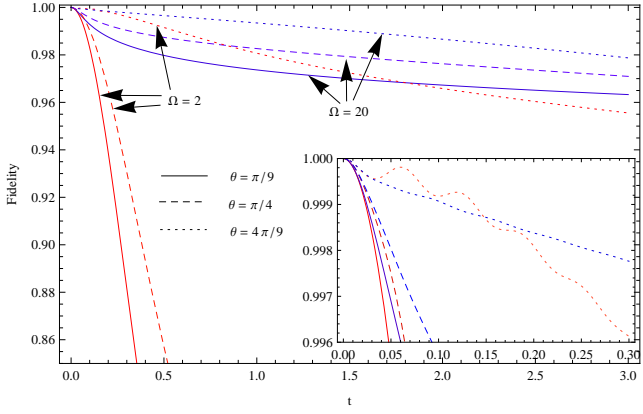}}
\caption{ The time-evolution of the fidelity $F(t)$ for the cases of Markovian $\Omega=20$ (blue) and non-Markovian $\Omega=2$ (red) environments.  The other parameter values are $B=100$, $\Omega_0=2$ and $T=0$. The inset enlarges the top left corner of the figure.} \label{fig1}
\end{figure}

In zero temperature and with $\Omega_0=2$ and $\int_{0}^{\infty}J(\omega)d\omega\equiv \lambda\Omega^{2}/2 =2$,  $F(t)$ is plotted in  Fig.~\ref{fig1} for non-Markovian ($\Omega=2$) and  Markovian ($\Omega=20$) cases.  It can be seen that the fidelity decreases more rapidly in the non-Markovian environment than in the Markovian environment.

The dependence of the fidelity on the azimuthal angle of the rotational magnetic field $\theta$ is clearly shown.  The inset in Fig.~\ref{fig1} shows the oscillation of the fidelity in the beginning if $\theta$ is large enough, which indicates feedback from the environment. After a certain period of time, the non-Markovian effect becomes  remarkable. The smaller the  value of $\theta$, the smaller the fidelity.

Now we consider the following evolution from the initial state  $\rho^{s}(0)$, which is designed
to cancel the dynamical phase and result in purely the  Berry phase~\cite{shi},
\begin{equation}
\rho^{s}(2T_{0})={\cal R}_{-\Omega_0} \Pi {\cal R}_{\Omega_0}\rho^{s}(0), \label{19}
\end{equation}
where ${\cal R}_{\Omega_0}$ represents an adiabatical rotation with constant angular frequency $\Omega_0$ for a time period $T_{0}=2\pi/\Omega_0$, which is followed by an a instantaneous $\pi$ pulse represented as  $\Pi \rho  \equiv A \rho A^{\dagger}$, where  $A\equiv|e(T_{0})\rangle\langle g(T_{0})|+|g(T_{0})\rangle\langle e(T_{0})|$) exchanges  $|e(T_{0})\rangle$ and $|g(T_{0})\rangle$. It is then followed by a reversed rotation represented as  ${\cal R}_{-\Omega_0}$. Consequently the Berry phase doubles while the dynamical phase cancels. The relative phase between $|e(2T_{0})\rangle$ and $|g(2T_{0})\rangle$ can be measured by state tomography. Note that the process consists of two parts, each of which is subject to a constant rotation, while the two parts are connected by an instantaneous reversal of  rotation is direction, with the final state of the first part being the initial state of the second part.

For an isolated system, under the adiabatic approximation, the final state is
 \begin{equation}
\begin{array}{l}
\rho^{s}_0(2T_{0})  =  \\
\frac{1}{2}\left(
       \begin{array}{cc}
        1+\sin\theta\cos4\Phi&-\cos\theta\cos4\Phi-i\sin4\Phi  \\
         -\cos\theta\cos4\Phi+i\sin4\Phi & 1-\sin\theta\cos4\Phi\\
       \end{array}
     \right),
     \end{array}
\end{equation}
where $\Phi=\pi (1-\cos\theta)$ is the Berry phase of $|g\rangle$.

A straightforward but lengthy calculation yields
$F(2T_{0})=Tr[\rho^{s}(2T_{0})\rho_0^{s}(2T_{0})]=
\frac{1}{2}(1+\cos4\Phi\sin\zeta_{2}+e^{-l_{1}-n_{1}-n_{2}}\cos4\Phi\sin\zeta_{2}(e^{n_{1}}(\cos \eta_{1}\cos\zeta_{1}\sin\zeta_{12}-e^{l_{1}} (1+\cos\zeta_{12}+2m_{2}))-e^{l_{1}}
\cos\zeta_{12}(\sin\zeta_{1}-1-2m_{1}))+\frac{1}{2}e^{-l_{1}-l_{2}-n_{1}}
(e^{n_{1}}\cos\zeta_{1}(\cos \eta_{1}\cos\zeta_{12}(\cos4\Phi\cos \eta_{2}\cos\zeta_{2}
-\sin4\Phi\sin \eta_{2})+(\cos \eta_{2}\sin4\Phi+\cos4\Phi\cos\zeta_{2}\sin \eta_{2})\sin \eta_{1})+e^{l_{1}}(\cos4\Phi\cos \eta_{2}\cos\zeta_{2}-\sin4\Phi\sin \eta_{2})\sin\zeta_{12}
(e^{n_{1}}-1+\sin\zeta_{1}-2m_{1}))$,
where $\zeta_{12}\equiv\zeta_{1}-\zeta_{2}$, $\eta_{1,2}\equiv T_{0}E_{1,2}+k_{1,2}(T_{0})$, where the subscript ``1'' or ``2" denotes the first or second cycle with angular frequency $\Omega_{0}$ or $-\Omega_{0}$, respectively.

\section{Pure Berry phase under adiabatic approximation  \label{berry} }

 In the adiabatic-limit i.e., $\zeta_{1}\approx  \zeta_{2}\approx 0$,   one obtains
$
F(2T_{0})=\frac{1}{2}[1+e^{-l_{1}(T_{0})-l_{2}(T_{0})}
\cos(4\Phi-T_{0}(E_{1}-E_{2})+(k_{1}(T_{0})-k_{2}(T_{0}))].$
 In $F(2T_0)$, the absence of energy dissipation terms $n_{1,2}(T_{0})$ and $m_{1,2}(T_{0})$  in the adiabatic limit  is  due to  the equal superposition of $|g\rangle$ and $|e\rangle$ in the initial state.

Thus the dephasing factors  and spectrum-induced phase shifts  have strong effects on  the Berry phase. The phase shifts $k_1$ and $k_2$  partially cancel each other, but the dephasing factors $l_1$ and $l_2$ add. In the adiabatic limit,  $E_{1,2} \approx B\mp\Omega_0\cos\theta$, we have $4\Phi -T_{0}(E_{1}-E_{2})=4\pi$, which goes away. The phase correction is thus
$
\delta\Phi =k_{1}(T_{0})-k_{2}(T_{0})\simeq4\Omega_0\sin^{2}\theta\cos\theta
\int_{0}^{T_{0}}\int_{0}^{t}s\cos(Bs)Re[\kappa(s)]dsdt,$
from which we note that $\delta\Phi$ is proportional to $\sin^{2}\theta\cos\theta$, no matter what the environment spectrum is.

The dephasing factor  is $
l_{1}(T_{0})+l_{2}(T_{0}) \simeq
\int_{0}^{T_{0}}\int_{0}^{t}(8\cos^{2}\theta+4\sin^{2}\theta\cos(Bs))Re[\kappa(s)]dsdt
\stackrel{t\rightarrow\infty}{\rightarrow} 2\sin^{2}\theta T_{0}\pi J(B)[2N(B)+1],$
where we have used  $\lim_{t\rightarrow\infty}
\frac{\sin((B-\omega)t)}{B-\omega}=\pi\delta(B-\omega)$.  Taking the limit $t\rightarrow\infty$ is equivalent to the Markovian approximation. The result   is consistent with the result in Ref.~\cite{whitney}, suggesting that the geometric nature of the phase correction and dephasing is model-insensitive.  We would like to emphasize that in the adiabatic limit, there are two major factors degrading  the Berry phase, that is, the dephasing effect and the environment-induced phase shift, which are intertwined together.

\begin{figure}
\scalebox{0.7}[0.7]{\includegraphics{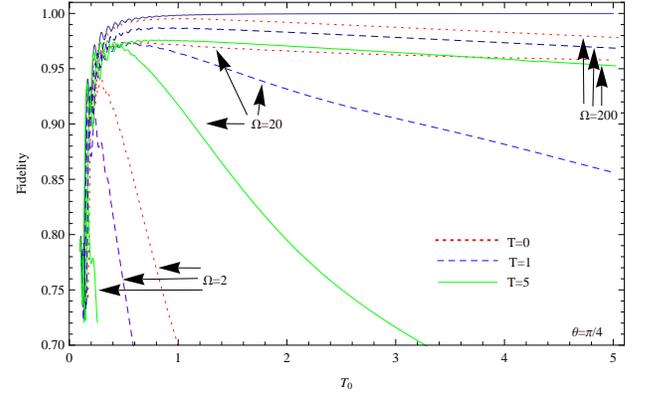}}
\caption{\label{fig2}  The fidelity of the final state $F(2T_{0})$ as a function of the time period $T_{0}$, for $\Omega=200, 20, 2$, and for   $T=0, 1, 5$. The other parameter values are $\theta=\pi/4$ and $B=100$. The uppermost solid line represent  the isolate system.    }
\end{figure}

\begin{figure}
\scalebox{0.85}[0.85]{\includegraphics{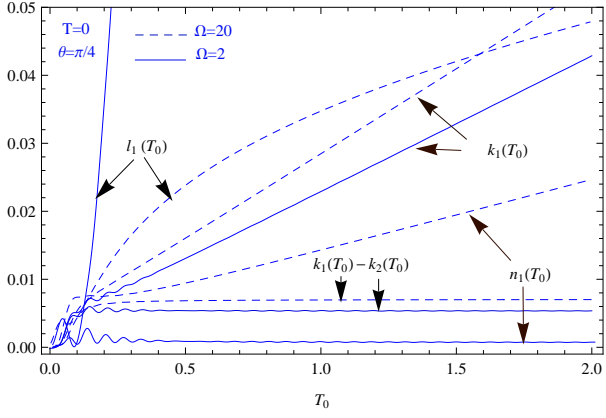}}
\caption{\label{fig3}$l_{1}(T_{0})$, $k_{1}(T_{0})$ and $k_{1}(T_{0})-k_{2}(T_{0})$ as functions of the time period $T_{0}$  for $\Omega=20, 2$.   We set $\theta=\pi/4$, $T=0$ and $B=100$.}
\end{figure}

\begin{figure}
\scalebox{0.95}[0.95]{\includegraphics{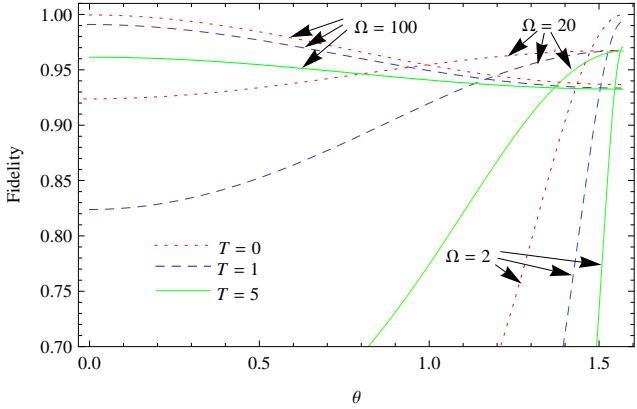}}
\caption{\label{fig4}  The dependence of fidelity $F(2T_{0})$ on  $\theta$ for different noise spectra and temperatures. We set $B=100$ and $\Omega_0=2$.}
\end{figure}

The dependence of the fidelity $F(2T_0)$  of the final state on the time period $T_0$ is shown in Fig.~\ref{fig2}, where the behavior of an isolated system is also shown for comparison. Note  that the  oscillation at small values of $T_{0}$, which exists in all cases including the isolated system,  is a manifestation of non-adiabatic error, which has been analyzed previously~\cite{shi0}.  In Fig.~\ref{fig2}, it can be seen that  in Markovian-limit and a low temperature environment, there is a wide range of $T_0$ where Berry phase  can be observed  as the fidelity is close to $1$. The non-Markovian spectral density is significant for low frequencies, where the effect of Berry phase is strongly degraded. This is similar to an experimental result~\cite{leek}.

We plot $n_{1}(T_{0})$, $l_{1}(T_{0})$, $k_{1}(T_{0})$, and $k_{1}(T_{0})-k_{2}(T_{0})$ together in Fig.~\ref{fig3}. $m_{1,2}(T_{0})$ is too close to  zero to be plotted.   In the adiabatic limit, the energy dissipation does not appear in the fidelity expression, while the Lamb shift is eliminated by the spin-echo technique, the dephasing effect plays a crucial role in degrading the coherence. Furthermore, an interesting characteristic exhibited in Fig.~\ref{fig4} is that the dependence of the fidelity $F(2T_{0})$   on the azimuthal angle $\theta$ varies with the noise spectrum. We note that in the expression of  $l(t)$,    the second term is negligible in the non-Markovian limit as it oscillates with frequency $B$, and thus $l(T_{0})\propto \cos^{2}\theta$. But in the Markovian limit,  $l(T_{0})\propto \sin^{2}\theta$. Therefore we can see that the fidelity of the final state at a certain temperature is a monotonically increasing function of the azimuthal angle $\theta$ in the non-Markovian case, but is a monotonically decreasing function of $\theta$ in the Markovian limit.

\section{More general situations \label{general} }

The above analysis can be extended to the situation that the magnetic field adiabatically travels along an arbitrary closed path: $\mathbf{B}(t)=B(\sin\theta(t)\cos\varphi(t),\sin\theta(t)\sin\varphi(t),\cos\theta(t))$, leading to a Berry phase  $\Phi=\frac{1}{2}\oint(1-\cos\theta)d\varphi$ for $|g(t)\rangle$, or $-\Phi$ for $|e(t)\rangle$. The transformed Hamiltonian   is $
\tilde{H}=U_{3}(t)(U_{2}(t)(U_{1}(t)H U_{1}^{\dag}(t)-iU_{1}(t)\dot{U}_{1}^{\dag}(t))U_{2}^{\dag}(t)
-iU_{2}(t)\dot{U}_{2}^{\dag}(t))U_{3}^{\dag}(t)-iU_{3}(t)\dot{U}_{3}^{\dag}(t)
=\tilde{H}'_{s}+\tilde{V}'+\tilde{H}_{e},$
where $U_{1}(t) = e^{\frac{i\varphi(t)}{2}\sigma_{z}}$, $U_{2}(t) = e^{\frac{i\alpha(t)}{2}\sigma_{y}}$, $U_{3}(t)= e^{\frac{i\beta(t)}{2}\sigma_{x}}$, $\tan\alpha(t) =\frac{B\sin\theta(t)}{B\cos\theta(t)-\dot{\varphi}(t)}$,
$\tan\beta(t)=\frac{\dot{\alpha}(t)}{\sqrt{(B \sin\theta(t))^{2}+(B\cos\theta(t)-\dot{\varphi}(t))^{2}}}$), $\tilde{H}'_{s}=\frac{1}{2}(E'(t)\sigma_{z}-\dot{\beta}(t)\sigma_{x})$ ($E'(t)=\sqrt{(B\sin\theta(t))^{2}+(B\cos\theta(t)-\dot{\varphi}(t))^{2}+\dot{\alpha}^{2}(t)}$), $\tilde{V}'=(\cos\alpha(t)\cos\beta(t)\sigma_{z}-\sin\alpha(t)\sigma_{x}+\cos\alpha(t)\sin\beta(t)\sigma_{y})\sum_{k}g_{k}(a^{\dag}_{k}+a_{k})$. To proceed, we treat  $\tilde{H}_{s}$ under first-order adiabatic approximation, while treat  $\tilde{V}'$ to zeroth-order  i.e., $
\tilde{H}'_{s}\approx\frac{B-\cos\theta(t)\dot{\varphi}(t)}{2}\sigma_{z},$
$\tilde{V}'\approx
(\cos\theta(t)\sigma_{z}-\sin\theta(t)\sigma_{x})\sum\limits_{k}g_{k}(a^{\dag}_{k}+a_{k})$. Then it can be obtained that    $n(t)$, $m(t)$, $l(t)$ and $k(t)$ replaced as $n'(t)$, $m'(t)$, $l'(t)$ and $k'(t)$ respectively, given by $n'(t)=4\int^{t}_{0}\int^{t_{1}}_{0}\sin\theta(t_{1})\sin\theta(t_{1}-t_{2})\cos(Bt_{2})Re[\kappa(t_{2})]dt_{2}dt_{1},$ $m'(t)=2\int^{t}_{0}\int^{t_{1}}_{0}\sin\theta(t_{1})\sin\theta(t_{1}-t_{2})Re[\tilde{\kappa}(t_{2})]
e^{n(t_{1})}dt_{2}dt_{1},$
$l'(t)=\int^{t}_{0}\int^{t_{1}}_{0}(4\cos\theta(t_{1})\cos\theta(t_{1}-t_{2})+2\sin\theta(t_{1})\sin\theta(t_{1}-t_{2})
\cos(Bt_{2}))Re[\kappa(t_{2})]dt_{2}dt_{1},$
$k'(t)=2\int^{t}_{0}\int^{t_{1}}_{0}\sin\theta(t_{1})\sin\theta(t_{1}-t_{2})\sin(Bt_{2})Re[\kappa(t_{2})]dt_{2}dt_{1}.$

\begin{figure}
\scalebox{0.7}[0.7]{\includegraphics{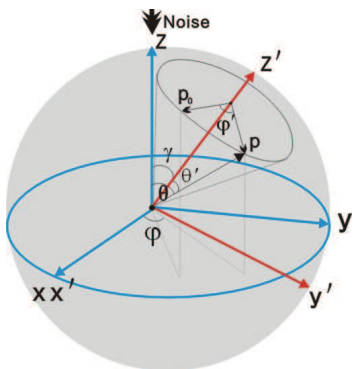}}
\caption{\label{fig5}(color on line) The path  traversed by the tip of the magnetic field (dot p). $\gamma$ is the angle between the $z$ axis and the $z'$ axis. $\varphi'=\omega_{0}t$ and $\theta'=const$. }
\end{figure}

\begin{figure}
\scalebox{1.1}[1.1]{\includegraphics{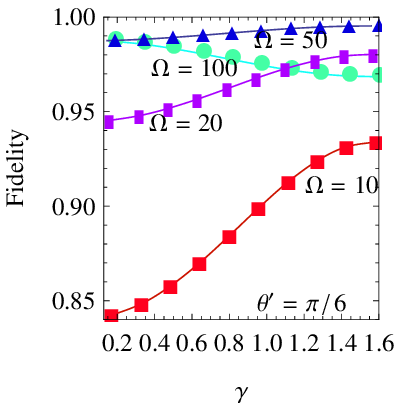}}\scalebox{1.1}[1.1]{\includegraphics{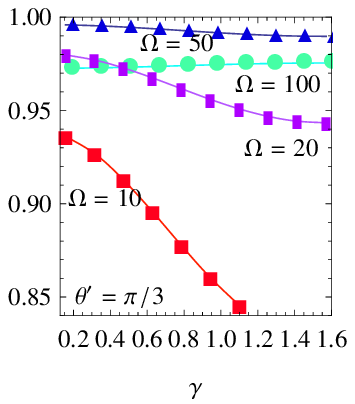}}
\caption{\label{fig6}(color on line) The Fidelity $F(2T_{0})$ as a function of the noise direction $\gamma$. We always set $B=100$, $T=0$, and $\Omega_{0}=2$.}
\end{figure}

As an example, consider the closed path traveled by the magnetic field is given by
\begin{equation}\label{Eq.18}
\left[
       \begin{matrix}
         \sin\theta\cos\varphi\\
          \sin\theta\sin\varphi\\
           \cos\theta\\
          \end{matrix}
     \right]=\left[
       \begin{matrix}
         1 & 0 & 0  \\
          0 & \cos\gamma & \sin\gamma  \\
         0 &  -\sin\gamma & \cos\gamma  \\
          \end{matrix}
     \right]
     \left[
       \begin{matrix}
         \sin\theta'\cos\varphi'\\
          sin\theta'\sin\varphi'\\
           \cos\theta'\\
          \end{matrix}
     \right],
\end{equation}
as shown in Fig.\ref{fig5}. Noting  $\theta_{2}(t)=\theta_{1}(T_{0}-t)$, we can investigate the response of the Berry phase for the noise direction defined by $\gamma$. The dependence on $\gamma$ indicates that for a certain noise environment we can always choose an optimal loop to minimize the decoherence effect. Fig.~\ref{fig6} also shows that in addition to the noise spectrum, the azimuthal angle $\theta'$ itself, as an intrinsic geometry parameter for the Berry phase, strongly affects the fidelity. some features of the fidelity $F(2T_0)$  for a smoothly rotating field persist in the present more general case.

We can also consider the multi-noise case based on Eq. (\ref{eq1}) by assuming that the frequency $\Omega_{0}$ is precisely controlled, this seems achievable in realistic devices, but an additional bosonic bath couples to the rotating component $B_{\|}$ in the $xy$ plane denoting the dipole interaction in the solid-qubit device \cite{leek}. In the rotating frame the Hamiltonian can be written as $\tilde{H}=\frac{E}{2}(\cos\alpha\sigma_{z}+
\sin\alpha\sigma_{x})+\sigma_{z}\sum\limits_{k}g_{zk}(a_{zk}^{\dag}+a_{zk})+
\sigma_{x}\sum\limits_{k}g_{xk}(a_{xk}^{\dag}+a_{xk})+
\sum\limits_{k}\omega_{zk}a_{zk}^{\dag}a_{zk}+
\sum\limits_{k}\omega_{xk}a_{xk}^{\dag}a_{xk}.$
Assuming the two independent baths possess the same autocorrelation function, i.e., $\langle \epsilon_{z}(t)\epsilon_{z}(s)\rangle=\langle \epsilon_{x}(t)\epsilon_{x}(s)\rangle$, we found that the decoherence no longer depends on the azimuthal angle $\theta$, and that  the spectrum-induced phase shift $\delta\Phi=0$.  The general feature of $F(2T_0)$ observed above is still valid now. However, the dephasing effect caused by low-frequency noise has been enhanced.

\section{SUMMARY \label{summary} }

To summarize, we have analyzed the effect of decoherence on the Berry phase by calculating the   fidelity between the reduced density matrix of the system coupled with an environment and the exact state of a closed system, both  starting from a same initial state.  This approach does not rely on any definition of the geometric phase in a mixed state. We use the master equation without any constraint on the correlation time of the bath, hence our discussions cover both the non-Markovian dynamics and Markovian limit. It is found that in the adiabatic limit, with a high frequency quantum noise, the deviation of the fidelity from $1$ is quite small, implying that the Berry phase is robust.  With a low frequency quantum noise, the fidelity is significantly lowered, implying that the Berry phase is significantly degraded. Our finding is beyond what can be obtained by making Markovian approximation, and is in accordance with the experimental result~\cite{leek}. We also note that the result is valid in the more general cases of an arbitrary path of the cycle and of the multi-noise. It is also noted that for the initial state considered, dephasing clearly dominates over  the energy dissipation and the Lamb shift. As the dephasing is path dependent, an optimal evolution loop can be chosen to protect the coherence. We hope that our analysis is useful for designing quantum gates based on geometric phases.
Finally, we note that it is interesting to combine the geometric phase approach with the dynamical decoupling approach~\cite{viola}, where it is also found that non-Markovian environmental noise causes phase randomization~\cite{uhrig}.

\acknowledgments
This work was supported by the National Science Foundation of China (Grant No. 11074048) and the Ministry of Science and Technology of China (Grant No. 2009CB929204).

\end{document}